\documentclass[aps,pre,twocolumn,superscriptaddress]{revtex4-1}

\usepackage{graphicx}
\usepackage{amsmath}

\usepackage[normalem]{ulem}

\usepackage[dvips]{color}

\begin{document}

\title{Local yield stress statistics in model amorphous solids}

\author{Armand Barbot}
\author{Matthias Lerbinger}
\author{Anier Hernandez-Garcia}
\author{Reinaldo	 Garc\'ia-Garc\'ia}
\affiliation{PMMH, ESPCI Paris/CNRS-UMR 7636/Univ.~Paris 6 UPMC/Univ.~Paris 7 Diderot, PSL Research Univ., 10 rue Vauquelin, 75231 Paris cedex 05, France}
\author{Michael L. Falk}
\affiliation{Departments of Materials Science and Engineering, Mechanical Engineering, and Physics and Astronomy, Johns Hopkins University, Baltimore, MD 21218}
\author{Damien Vandembroucq}
\author{Sylvain Patinet}
\affiliation{PMMH, ESPCI Paris/CNRS-UMR 7636/Univ.~Paris 6 UPMC/Univ.~Paris 7 Diderot, PSL Research Univ., 10 rue Vauquelin, 75231 Paris cedex 05, France}
\email[]{sylvain.patinet@espci.fr}

\date{\today}

\begin{abstract}
We develop and extend a method presented in [S. Patinet, D. Vandembroucq, and M. L. Falk, Phys.
Rev. Lett., 117, 045501 (2016)] to compute the local yield stresses at
the atomic scale in model two-dimensional Lennard-Jones glasses
produced via differing quench protocols. This technique allows us to
sample the plastic rearrangements in a non-perturbative manner for
different loading directions on a well-controlled length
scale. Plastic activity upon shearing correlates strongly with the
locations of low yield stresses in the quenched states. This
correlation is higher in more structurally relaxed systems.  The
distribution of local yield stresses is also shown to strongly depend
on the quench protocol: the more relaxed the glass, the higher the
local plastic thresholds. Analysis of the magnitude of local plastic
relaxations reveals that stress drops follow exponential
distributions, justifying the hypothesis of an average characteristic
amplitude often conjectured in mesoscopic or continuum models. The
amplitude of the local plastic rearrangements increases on average
with the yield stress, regardless of the system preparation. The local
yield stress varies with the shear orientation tested and strongly
correlates with the plastic rearrangement locations when the system is
sheared correspondingly. It is thus argued that plastic rearrangements
are the consequence of shear transformation zones encoded in the glass
structure that possess weak slip planes along different
orientations. Finally, we justify the length scale employed in this
work and extract the yield threshold statistics as a function of the
size of the probing zones. This method makes it possible to derive
physically grounded models of plasticity for amorphous materials by
directly revealing the relevant details of the shear transformation
zones that mediate this process.
\end{abstract}


\maketitle

\section{\label{sec:introduction} Introduction}

Despite numerous advances during the last two decades, a physical
description of plasticity in amorphous materials, known to be
quantitatively tied to well-characterized atomistic processes, remains
a grand
challenge~\cite{procaccia_physics_2009,barrat_heterogeneities_2011,rodney_modeling_2011}. All
the constitutive laws describing the plastic flow of this large class
of materials, such as glasses, amorphous polymers or gels, remain based on
phenomenological assumptions. This fact is due to the lack of
systematic characterization of elementary mechanisms of plasticity at
the atomic scale. For the amorphous solids, the absence of
  crystalline order prevents, by definition, any identification of
  crystallographic defects such as dislocations. These defects are,
however, the quanta of plastic deformation from which it has been
possible to derive constitutive equations of crystal plasticity on
robust physical grounds. In amorphous materials, plastic deformation
manifests as local rearrangements~\cite{argon_1979} exhibiting a
  broad distribution of sizes and shapes~\cite{lerner_locality_2009},
  nonaffine
    displacements~\cite{zaccone_microscopic_2014,laurati_long-lived_2017}
    and connectivity changes between
    particles~\cite{van_doorn_linking_2017} that lead to a
redistribution of elastic stresses in the
system~\cite{chattoraj_elastic_2013,lemaitre_structural_2014}. By
analogy with dislocations, it therefore appears natural to try to
describe the plastic flow from the dynamics of localized “defects”
commonly referred to as Shear Transformation Zones
(STZs)~\cite{falk_dynamics_1998}.

A variety of metrics have been proposed to locate and characterize the
defects that control plastic activity including structural properties
(free volume~\cite{spaepen_microscopic_1977}, packing
\cite{jack_information-theoretic_2014}, short-range order
\cite{shi_stress-induced_2007} and internal stress
\cite{tsamados_study_2008}) and linear responses measures (elastic
moduli~\cite{tsamados_local_2009} and localized soft vibrational modes~\cite{widmer-cooper_irreversible_2008,widmer-cooper_localized_2009,tanguy_vibrational_2010,ghosh_connecting_2011,manning_vibrational_2011,chen_measurement_2011,mosayebi_soft_2014,ding_soft_2014,schoenholz_understanding_2014}). Unfortunately,
the definition of these structural properties are often
system-dependent and have shown a relatively low predictive power with
respect to the plastic activity~\cite{tsamados_local_2009,jack_information-theoretic_2014}. The
  approaches based on linear response measures (e.g. soft vibrational
  mode analysis) have shown that the correlation between these local
  properties and the location of plastic rearrangements decreases
  rather quickly as the system is deformed plastically since they are
  derived from perturbative calculations~\cite{patinet_connecting_2016}.

To address these problems, new methods have recently been
proposed. They are based either on combinations of static and dynamic
properties (atomic volume and vibrations)~\cite{ding_universal_2016},
on the non-linear plastic
modes~\cite{lerner_micromechanics_2016,zylberg_local_2017} or on
machine learning
methods~\cite{cubuk_identifying_2015,cubuk_structural_2016,schoenholz_structural_2016}. These
approaches allow the calculation of local fields (respectively named
by their authors \textit{flexibility volume}, \textit{local thermal
  energy} and \textit{softness field}) which show abilities to detect
plastic defects far superior to previous attempts. Independently of
their high degree of correlation, they nevertheless have the
disadvantage of giving access only to quantities that are not directly
related to local yield criteria that are more commonly used
  in models of plasticity. Moreover, with the exception of works
  based on the vibrational soft
  modes~\cite{rottler_predicting_2014,smessaert_structural_2014}, the
  vast majority of these previous approaches only gives access to scalar
  quantities, which by definition also neglect the important
  orientational aspect inherent in STZ activity.

It is precisely in this context that we have recently developed a numerical technique to systematically measure the local yield stress field of an amorphous solid on the atomic scale~\cite{patinet_connecting_2016}. This new approach responds to some issues raised previously by providing access to a natural quantity in the context of plasticity, i.e. a local yield stress, and shows an extremely strong correlation with the plastic activity. In addition, this method is non-perturbative and can investigate large strains. It also gives access to a tensorial quantity and is thus able to describe several possible directions of flow. It is therefore an ideal candidate to quantitatively characterize the relationship between structure and plasticity.

In this paper, we present in detail the principles of this method. The statistics of local yield stress are calculated in a model glass synthesized from different quench protocols. The correlation between local slip threshold and plastic activity is investigated as a function of the degree of relaxation of the system. The method is subsequently extended to the study of the amplitudes of the plastic relaxations. Additionally, the consequences of the orientation of the mechanical loading are examined. Finally, we address the effect of the length scale over which the local yield stress field is computed.

\section{\label{sec:methods} Simulation methods}

\subsection{\label{sec:preparation} Sample preparation}
We performed molecular dynamics and statics simulations with the LAMMPS open software~\cite{plimpton_fast_1995}. The object of study is a two-dimensional binary glass which is known for its good glass formability~\cite{lancon_structural_1984,widom_quasicrystal_1987}. It has previously been used to study the plasticity of amorphous materials~\cite{falk_dynamics_1998,shi_structural_2005,shi_strain_2005,shi_evaluation_2007}. One hundred samples each containing $10^4$ atoms were obtained by quenching liquids at constant volume. The density of the system is kept constant and equals $10^4/(98.8045)^2 \approx 1.02$. As in~\cite{shi_strain_2005}, we choose our composition such that the number ratio of large (L) and small (S) particles equals $N_{L}:N_{S}=(1+\sqrt{5})/4$. The two-types of atoms interact via standard  $6-12$ Lennard-Jones interatomic potentials. In the following, all units will therefore be expressed in terms of the mass $m$ and the two parameters describing the energy and length scales of interspecies interaction, $\epsilon$ and $\sigma$, respectively. Accordingly, time will be measured in units of $t_{0}=\sigma\sqrt{m/\epsilon}$. In the present study, these potentials have been slightly modified to be twice continuously differentiable functions. This is done by replacing the Lennard-Jones expression for interatomic distances greater than $R_{in}=2\sigma$ by a smooth quartic function vanishing at a cutoff distance $R_{cut}=2.5\sigma$. For two atoms $i$ and $j$ separated by a distance $r_{ij}$:
\begin{equation}
\label{eq:interatomicpotential}
U(r_{ij}) =
\begin{cases}
4\epsilon \left[ \left(\frac{\sigma}{r_{ij}}\right)^{12}-\left(\frac{\sigma}{r_{ij}}\right)^{6}\right]+A, \quad \text{  for } r_{ij}<R_{in}\\
\sum_{k=0}^{4}C_{k}(r_{ij}-R_{in})^{k}, \quad \text{for } R_{in}<r_{ij}<R_{cut}\\
0, \qquad \qquad \qquad \qquad \qquad \qquad \ \ \text{for } r_{ij}>R_{cut},
\end{cases}
\end{equation}
with
\begin{subequations}
\label{eq:interatomicpotentialcoeff}
\begin{align*}
A&=C_{0}-4\epsilon\left[\left(\frac{\sigma}{R_{in}}\right)^{12}-\left(\frac{\sigma}{R_{in}}\right)^{6}\right]\\
C_{0}&=-(R_{cut}-R_{in})(3C_{1}+C_{2}(R_{cut}-R_{in}))/6 \\
C_{1}&=24\epsilon\sigma^6(R_{in}^6-2\sigma^6)/R_{in}^{13} \\
C_{2}&=12\epsilon\sigma^6(26\sigma^6-7R_{in}^6)/R_{in}^{14} \\
C_{3}&=-(3C_{1}+4C_{2}(R_{cut}-R_{in}))/(3(R_{cut}-R_{in})^2) \\
C_{4}&=(C_{1}+C_{2}(R_{cut}-R_{in}))/(2(R_{cut}-R_{in})^3). \\
\end{align*}
\end{subequations}
\begin{figure}
\textbf{(a)}
\includegraphics[width=0.85\columnwidth]{stress_vs_strain}\\
\textbf{(b)}
\includegraphics[width=0.85\columnwidth]{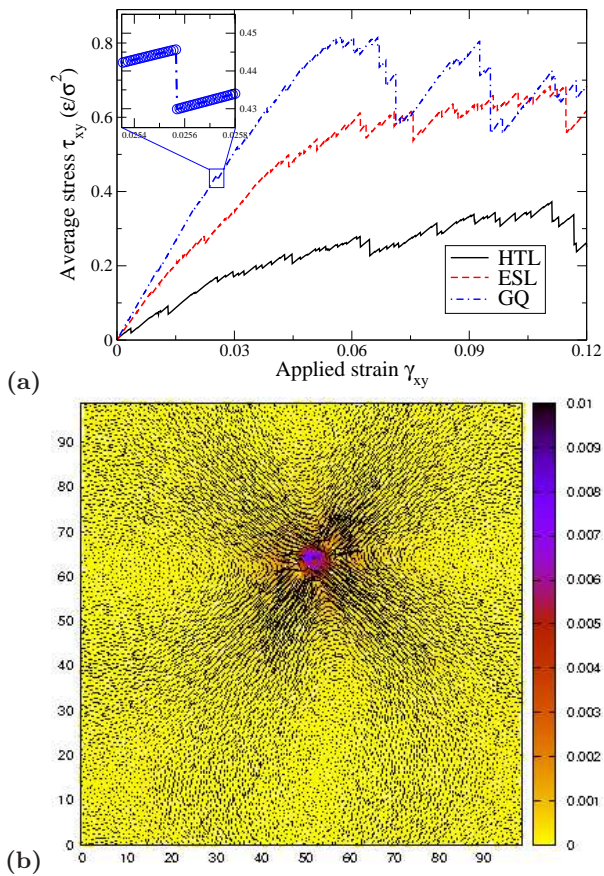}
\caption{\label{fig:mesoscopic} a) Typical stress-strain curves for the three quench protocols: instantaneous quench from a High Temperature Liquid (HTL, continuous line), instantaneous quench from an Equilibrated Supercooled Liquid (ESL, dashed line) and a Gradual Quench (GQ, dash-dotted line). The inset shows a zoom of a stress drop corresponding to one plastic event. b) Plastic event computed between the onset of instability and just after the event: the arrows and the color scale are the displacement $\vec{u}$ and maximum shear strain $\eta_{VM}$ fields, respectively. For the sake of clarity, the arrows are magnified by a factor $200$ and deleted in the core region. The atom with the maximum shear strain gives the location of the plastic rearrangement.}
\end{figure}
Periodic boundary conditions are imposed on square boxes of linear
dimensions $L=98.8045 \sigma$. For the non-smoothed version of the
  interatomic potential, the glass transition temperature $T_{g}$ of
  this system is known to be approximately $T_{g}=0.325 \epsilon/k$,
  where $k$ is the Boltzmann constant~\cite{shi_strain_2005}. This
  temperature corresponds to the mode coupling temperature which is an
  upper bound of the glass transition
  temperature~\cite{shi_structural_2005}. In order to highlight the
links between the microstructure, the stability of glasses and their
mechanical properties, three different quench protocols are
considered. The first two kinds of glass are obtained after
instantaneous quenches from High Temperature Liquid (HTL) and
Equilibrated Supercooled Liquid (ESL) states at $T=9.18 T_{g}$ and
$T=1.08 T_{g}$, respectively. The last protocol consists in a Gradual
Quench (GQ) in which temperature is continuously decreased from a
liquid state, equilibrated at $1.08 T_{g}$, to a low-temperature solid
state at $0.092 T_{g}$, over a period of $10^{6} t_{0}$ using a
Nose-Hoover
thermostat~\cite{nose_unified_1984,hoover_canonical_1985}. Afterwards,
the system is quenched instantaneously as well. All quench protocols
are followed by a static relaxation via a conjugate gradient method to
equilibrate the system mechanically at zero temperature. The forces on
each atom are minimized up to machine precision. The same relaxation
algorithm is used hereafter to study the response to mechanical
loading.

This approach produces three highly contrasting types of amorphous solids. The greater the temperature from which the system has fallen out of equilibrium, the less relaxed the system~\cite{sastry_signatures_1998,debenedetti_supercooled_2001}. This fact is clearly reflected in the values of the average potential energies per atom of the generated inherent states, equal to $-2.1015\pm0.0011$, $-2.3248\pm 0.0015$ and $-2.3977\pm0.0019 \epsilon$ for the HTL, ESL and GQ protocols, respectively.

\subsection{\label{sec:loading} Mechanical loading: generation of plastic events}

Beginning from a quenched unstrained configuration, the glasses are deformed in simple shear imposing Lees-Edwards boundary conditions with an Athermal Quasi Static method (AQS)~\cite{malandro_molecular-level_1998,malandro_relationships_1999,maloney_universal_2004,maloney_amorphous_2006}
. We apply a series of deformation increments $\Delta\gamma_{xy}$ to the material by moving the atom positions $\vec{r}$ following an affine displacement field such that $r_{x} \rightarrow r_{x}+r_{y}\Delta\gamma_{xy}$ and $r_{y} \rightarrow r_{y}$. After each deformation increment, we relax the system to its mechanical equilibrium. 

In order not to miss plastic events, a sufficiently small strain
increment equal to $\Delta\gamma_{xy}=10^{-5}$ is chosen. Plastic
events are detected when the computed stress $\tau_{xy}$ decreases, a
signature of mechanical instability. A reverse step
$-\Delta\gamma_{xy}$ is systematically applied after each stress drop
to confirm that the strains generated in the solid are irreversible
when the stress criterion is satisfied. The observed response is
typical for amorphous materials and is characterized by reversible
elastic branches interspersed by plastic events as illustrated in
Fig.~\ref{fig:mesoscopic}a. The more relaxed the system, the stiffer
and harder the glass. In agreement with~\cite{shi_strain_2005}, we
  observe that the localization of plastic strain, under load,
  increases with the degree of relaxation of the initial state. Note
that in the case of the GQ system a strong localization of the strain
is observed around $\gamma_{xy}\sim0.06$ due to shear banding.

The average shear moduli $\mu$ are obtained from the ratio between the stress response of the entire system following a deformation increment $\Delta\tau_{xy}/\Delta\gamma_{xy}$, i.e. the slope at the origin of the stress-strain curves reported in Fig.~\ref{fig:mesoscopic}a. $\mu$ equals $8.85$, $14.73$ and $19.03$ for HTL, ESL and GQ systems, respectively. As expected, the stiffness of the system increases with its level of relaxation.

\subsection{\label{sec:strain_field} Strain field computation}

In order to quantify the correlation between deformation thresholds and plastic rearrangements two types of deformation fields are calculated. The first corresponds to the plastic deformation induced by a single plastic event, the second describes the total cumulative deformation. The displacement field of the former is calculated using the difference between the position of atoms after and just before that instability occurs minus the applied affine displacement increment. The displacement field of the latter is merely computed as the difference between the position of atoms in configurations at a given strain and the as-quenched state, that is to say the state that has not yet been deformed mechanically. The Green-Lagrange strain tensor $\eta_{ij}$ is then evaluated from displacement fields $\vec{u}$ of each atom following the coarse-graining method developed in~\cite{hinkle_coarse_2017} based on the atomic gradient tensor evaluation~\cite{zimmerman_deformation_2009}. An octic polynomial coarse-graining function $\phi(r)$ is employed~\cite{lemaitre_structural_2014}. This function has a single maximum and continuously vanishes at $r=R_{CG}$ where $r$ is the distance between the strain evaluation location and the atom positions. It is expressed as:
\begin{equation}
\label{eq:coarse-grainingfunction}
\phi(r) =
\begin{cases}
\frac{15}{8\pi R_{CG}^2}(1-2(\frac{r}{R_{CG}})^4+(\frac{r}{R_{CG}})^8), &\text{for } r<R_{coars}\\
0, &\text{otherwise}.
\end{cases}
\end{equation}
It is desirable to consider a large enough coarse-graining length scale so that continuum mechanics quantities make sense while keeping it as small as possible in order to account for heterogeneity and to preserve spatial resolution. To this aim, we choose $R_{CG}=5\sigma$. On this scale, a continuous description makes sense (Hooke's law holds) but the solid is still anisotropic and heterogeneous~\cite{goldhirsch_microscopic_2002,goldenberg_particle_2007,tsamados_local_2009}.

To simplify the analysis, we choose to work with a scalar quantity by computing the maximum of the shear deformation $\eta_{VM}=\sqrt{((\eta_{xx}-\eta_{yy})/2)^2+\eta_{xy}^2}$. The positions of a plastic rearrangement are then defined as the position of the atom having undergone the maximum $\eta_{VM}$ during a plastic event. This approach allows us to obtain the successive positions of localized plastic rearrangements during deformation from the quenched state as exemplified in Fig.~\ref{fig:mesoscopic}b.

\begin{figure}
\textbf{(a)}
\includegraphics[width=0.75\columnwidth]{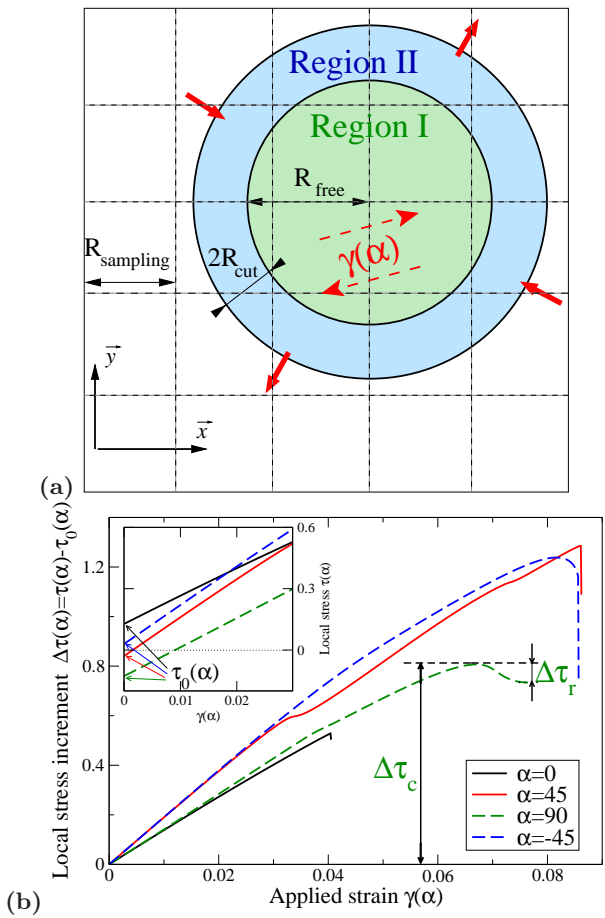}\\
\textbf{(b)}
\includegraphics[width=0.85\columnwidth]{stress_vs_strain_local}
\caption{\label{fig:local_method} a) Schematic drawing of the local yield stress computation on a regular square grid of mesh size $R_{sampling}$. Region I (of radius $R_{free}$) is fully relaxed while region II (of width $2R_{cut}$) is forced to deform following an affine pure shear deformation in the $\alpha$ direction. b) Typical local stress increment-strain curves of the Region I for different loading directions $\alpha=0$, $45$, $90$ and $-45^{\circ}$. The measurement of the threshold $\Delta\tau_c$ and relaxation $\Delta\tau_r$ are represented for $\alpha=90^{\circ}$. Inset: zoom one the low strain region of the corresponding stress-strain curves, i.e. without subtracting the initial local shear stress within the as-quenched glass $\tau_{0}(\alpha)$.}
\end{figure}

\section{\label{sec:local_method} Local yield stress measurement method}

We used a method developed in~\cite{patinet_connecting_2016} which allows us to sample the local flow stresses of glassy solids for different loading directions. Similar techniques have been employed to sample the local elastic moduli~\cite{mizuno_measuring_2013} or the yield stresses along a single direction in model glasses~\cite{sollichtalks_2011,Puosi_Probing_2015}. The principle of the numerical method is illustrated in Fig.~\ref{fig:local_method}a. It consists in locally probing the mechanical response within an embedded region of size $R_{free}$  (named region I) by constraining the atoms outside of it (named region II) to deform in a purely affine manner. Only the atoms within the region I are relaxed and can deform nonaffinely. Plastic rearrangements are, thus, forced to occur within this region and the local yield stress can be identified.

In~\cite{patinet_connecting_2016}, the embedded region I was centered on every atom of the system to test the reliability of the method. Here, to lay the groundwork for an up-scaling strategy, we rather sample the local yield stress on a regular square grid of size $R_{sampling}$. Furthermore, the non-interacting atoms, located farther than a distance $R_{free}+2R_{cut}$ from the center of the probed region, are deleted during the local loading simulations, thereby speeding up the computation. Unless mentioned explicitly, we chose $R_{sampling}=L/39 \approx R_{cut}$ and $R_{free}=5\sigma$, consistently with the coarse-graining scale $R_{CG}$ used for the strain computation.
The yield criterion and the AQS incremental method are the same as those described in section \ref{sec:loading} to shear the system remotely. At this scale the amorphous system is highly heterogeneous and the yield stress may not be the same for all orientations of the imposed shear. We thus sample the mechanical response using pure shear loading conditions in different loading directions $\alpha$. Eighteen directions, uniformly distributed between $\alpha=-90^{\circ}$ and $\alpha=90^{\circ}$, are investigated. $\alpha=0^{\circ}$ corresponds to the remote simple shear direction $\alpha_l$. 

\begin{figure*}
\includegraphics[width=1.9\columnwidth]{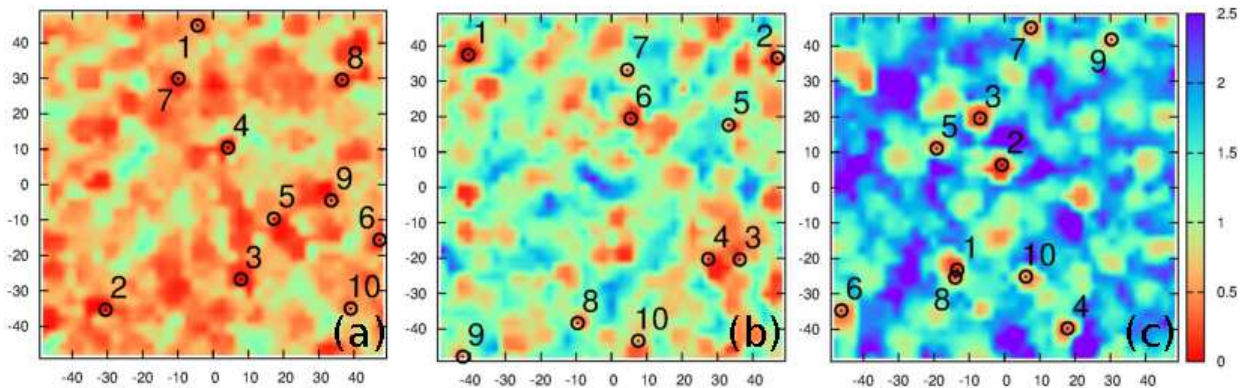}
\caption{\label{fig:yield_stress_map_quench_effect} Local yield stress maps computed on a regular grid for the three different quench protocols: a) HTL, b) ESL and c) GQ. The first ten plastic event locations are shown as open black symbols numbered by order of appearance during remote shear loading. Plastic events clearly tend to occur in regions characterized by low yield stresses.}
\end{figure*}

To save computational time, the strain increment is this time equal to $\Delta\gamma=10^{-4}$. The shear stress in the $\alpha$ direction $\tau(\alpha)$ is computed over the atoms that belong to region I using the Irving and Kirkwood formula~\cite{irving_statistical_1950} as a function of the applied strain. The local region is sheared up to the first mechanical instability occurring at a critical stress $\tau_c(\alpha)$. Even at rest, the glasses feature non-zero internal stress due to the frustration inherent to amorphous solids (see inset of Fig.~\ref{fig:local_method}b). A more relevant quantity to link the local properties with plastic activity is thus the amount of stress needed to trigger a plastic rearrangement. The initial local shear stress state within the as-quenched glass $\tau_0(\alpha)$ is thus subtracted from the critical stress to get the local stress increase that would trigger an instability $\Delta\tau_c(\alpha)=\tau_c(\alpha)-\tau_0(\alpha)$. Local shear stress-strain curves are exemplified for four different directions in Fig.~\ref{fig:local_method}b. It can be observed that the mechanical response depends on the loading orientation. As expected from elasticity theory, we verify that $\tau_{0}(\alpha)=-\tau_{0}(\alpha+90^{\circ})$ and that the local shear moduli $\mu(\alpha)=\mu(\alpha+90^{\circ})$ as shown in the inset of Fig.~\ref{fig:local_method}b \footnote{These symmetries are related to the rotation of the stress tensor and the elasticity tensor in the $xy$-plane through an angle $\alpha$ about the center of the patch. Noting the new coordinates $(x',y')$ of a point $(x,y)$ after rotation, the shear stress along $\alpha$ is equal to $\tau(\alpha)=\tau_{x'y'}=(\sigma_{yy}-\sigma_{xx})\sin(2\alpha)/2+\tau_{xy}\cos(2\alpha)$ and thus $\tau(\alpha)=-\tau(\alpha+90^{\circ})$. The shear modulus along $\alpha$ is equal to $\mu(\alpha)=C_{x'y'x'y'}=(C_{xxxx}+C_{yyyy}-2C_{xxyy})\sin(2\alpha)^2/4+(C_{xxxy}-C_{yyxy})\sin(4\alpha)/2+C_{xyxy}\cos(2\alpha)^2$ and thus $\mu(\alpha)=\mu(\alpha+90^{\circ})$.}. On the other hand, the critical stress increments $\Delta\tau_{c}$ do not show elastic symmetry and depend on the orientation considered. The computation of $\Delta\tau_{c}$ is repeated systematically for all the grid points of the system and for the different loading directions. 

We now want to consider the implications of the field of local
$\Delta\tau_c(\alpha)$ for a particular direction of remote loading
$\alpha_l$. For this purpose, we make the simplifying assumption of
homogeneous elasticity within the system or, equivalently, of
localization tensor equal to the identity tensor. Of course, the
  elasticity in this system at this length scale is heterogeneous and
  leads to non-affine displacements under remote loading as shown
  in~\cite{tsamados_local_2009}. This assumption is simply used to be
  able to estimate the stress felt by a local zone due to a
  remote loading. This assumption will be discussed further in
section \ref{sec:orientation_effects}. If the applied shear stress is
homogeneous in the glass, plastic rearrangement that would be
activated for a given site is the minimum (positive)
$\Delta\tau_c(\alpha)$ projected along the remote loading
direction. This may be expressed as:
\begin{equation}
\label{eq:loctauy}
\Delta\tau_{y}=\min_{\alpha}\frac{\Delta\tau_{c}(\alpha)}{\cos(2[\alpha-\alpha_l])} ~~ \textrm{with} ~~ |\alpha-\alpha_l| < 45^{\circ}.
\end{equation}
Maps of local $\Delta\tau_{y}$ are shown in the Fig.~\ref{fig:yield_stress_map_quench_effect} for the three quench protocols. One distinguishes a correlation length corresponding to the size of region I. Indeed, the same shear transformation zone can be activated for several grid locations if its threshold is smaller than others in its vicinity which leads to the assignment of close $\Delta\tau_{y}$ values on a scale $\sim R_{free}$. The influence of the size of region I will be addressed in section \ref{sec:length_scale}.

\section{\label{sec:local_rearrangement_statistics} Local rearrangement statistics}

\subsection{\label{sec:distributions_of_local_yield_stress} Distributions of local yield stress}

\begin{figure}
\includegraphics[width=0.9\columnwidth]{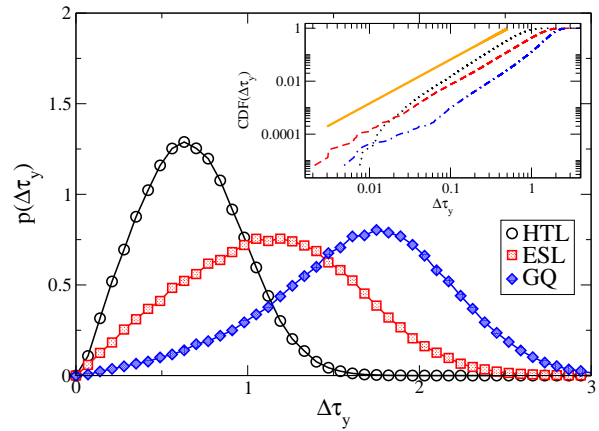}
\caption{\label{fig:PDF_yield_stress} Probability distribution function of the local yield stresses for the three different quench protocols. The corresponding cumulative distribution functions are represented in log-log scale in the inset. The straight line is a power law of exponent $1+\theta=1.6$, i.e. the expected scaling of the integral of the probability distribution function as $\Delta\tau_{y}$ approaches zero.}
\end{figure}

The effect of the quench protocol on the yield stress maps shown in Fig.~\ref{fig:yield_stress_map_quench_effect} is remarkable. It is readily apparent that the lower the temperature at which the system falls out of equilibrium during its synthesis, the more mechanically stable the glass. An advantage of our method is that it allows one to assess stability not from the global scale, as in Fig.~\ref{fig:mesoscopic}a, but locally. The HTL system shows an overabundance of small energy barriers characteristic of systems far from equilibrium. In contrast, the GQ system has a low proportion of soft zones embedded in a hard skeleton~\cite{shi_strain_2005,shi_does_2006}. As expected, ESL presents an intermediate situation. More quantitatively, the distributions of $\Delta\tau_{y}$ are computed for the three quenching protocols as shown in Fig.~\ref{fig:PDF_yield_stress}. The probability densities $p(\Delta\tau_{y})$ are noticeably shifted toward higher values with increasing system stability, weak areas being depopulated.

Through this method we are able to analyse the statistics of the sites
that are about to rearrange plastically. Previous work (based on
mean-field theoretical approaches~\cite{lin_mean_field_2016},
atomistic
simulations~\cite{karmakar_statistical_2010,hentschel_stochastic_2015}
and mesoscopic
simulations~\cite{lin_density_2014,lin_criticality_2015}) proposed a
scaling for these soft areas such as $\lim_{\Delta\tau_{y} \to 0}
P(\Delta\tau_{y})\sim\Delta\tau_{y}^{\theta}$ where $\theta$ is a
non-trivial exponent. For systems at rest, i.e. after the quench, it
was shown that $\theta\approx
  0.6$~\cite{karmakar_statistical_2010,hentschel_stochastic_2015}. From
detailed inspections of our results, as shown as inset of
Fig.~\ref{fig:PDF_yield_stress}, it seems difficult to extract this
exponent with the exception of ESL. In any case, it appears that the
behavior of $P(\Delta\tau_{y})$ in the limit of zero $\Delta\tau_{y}$
varies with the preparation of the system. In the case of the GQ
system, a smaller exponent can even be observed while for HTL a larger
one can be fitted for small threshold values. Several reasons may
account for this disagreement, including the lack of statistics or the
size of the strain increment $\Delta \gamma$. The fixed boundary
conditions applied during local probing may also prevent some
relaxation of the system. Also, notably, the previous approaches have
considered the distribution of critical strains applied to the whole
system which is strictly equivalent to our approach for an elastically
homogeneous glass, a strong assumption at this length
scale~\cite{tsamados_local_2009}. The answer to this question deserves
more investigation, which is outside the scope of the present study.

\subsection{\label{sec:correlation_with_plastic_activity} Correlation with plastic activity}

The position of the first ten plastic rearrangements during remote loading are illustrated in Fig.~\ref{fig:yield_stress_map_quench_effect}. Plastic rearrangements clearly tend to occur in the soft zones, i.e. for areas in which $\Delta\tau_y$ are small. To quantify this correlation, we apply the same method as in~\cite{patinet_connecting_2016}. 

\begin{figure}
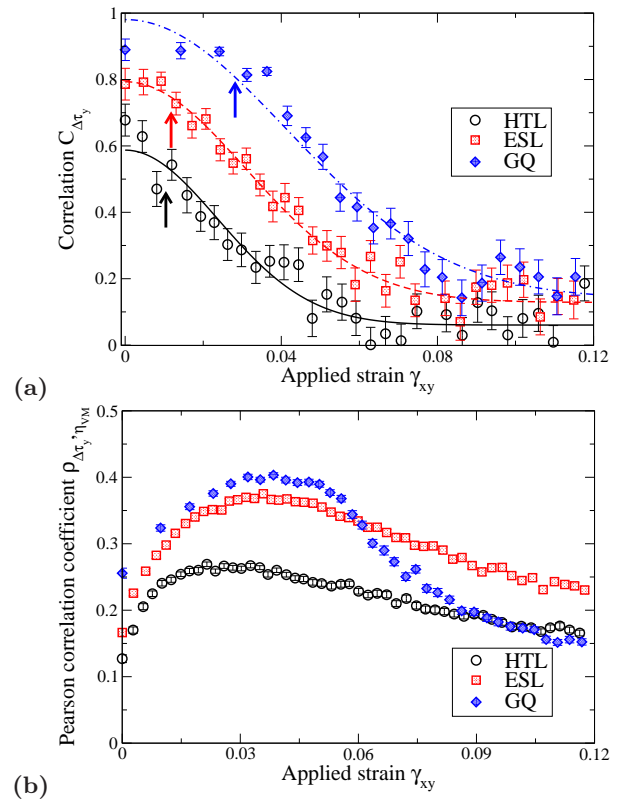

\textbf{(a)}
\includegraphics[width=0.85\columnwidth]{Correlation_CDF_QP_vs_Strain}\\
\textbf{(b)}
\includegraphics[width=0.85\columnwidth]{Correlation_Pearson_QP_vs_Strain}
\caption{\label{fig:correlations} Correlation between the local yield stresses computed in the quenched state and the locations of the plastic rearrangement as a function of the applied strain for the three quench protocols. The error bars correspond to one standard deviation. a) Correlation computed from individual plastic rearrangements using Eq.~(\ref{eq:correlation_CDF}). The arrows correspond to the average strain of the tenth plastic event. The lines are empirical $A+Be^{-(\gamma/\gamma_{d})^2}$ fits from which the decorrelation strain $\gamma_{d}$ is estimated. b) Pearson correlation coefficient computed between the local yield stress fields and the local strain fields using Eq.~(\ref{eq:correlation_Pearson}). Error bars are smaller than symbols.}
\end{figure}

We propose a correlation coefficient that allows us to relate the order of appearance of zones in which the plastic arrangements appear and a local scalar field, the local yield stress. The aim here is to compute the predictive power of a structural indicator for the location of successive rearrangements from the sole knowledge of the initial state of the system, i.e. before deformation. To achieve this, the correlation coefficient is computed from the value of the cumulative distribution function of $\Delta\tau_y$ corresponding to the point of the grid $i_{max}$, i.e. the closest to the location of the plastic rearrangement (determined according to the method described in the Sec. \ref{sec:loading}). The correlation coefficient is defined as:
\begin{equation}
\label{eq:correlation_CDF}
C_{\Delta\tau_{y}}=1-2\overline{CDF}[\Delta\tau_{y}(i_{max},\gamma_{xy})],
\end{equation}
where $\overline{CDF}$ is the disorder average of the cumulative
distribution function. $C_{\Delta\tau_{y}} \sim 1$ indicates a perfect
correlation, i.e. a localized plastic rearrangement on the lowest
yield threshold grid point ($CDF=0$), while $C_{\Delta\tau_{y}} \sim
0$ means an absence of correlation. $C_{\Delta\tau_{y}}$ is calculated
for all plastic events as a function of the deformation applied
$\gamma_{xy}$ as shown in Fig.~\ref{fig:correlations}a. Note that
  relation (\ref{eq:correlation_CDF}) neglects the stress
  redistribution due to successive rearrangements. Moreover, it only
  takes into account the rearrangements producing the maximum of local
  shear strain located at $i_{max}$ and therefore ignores the
  possibility that a plastic event may be composed of several
  localized rearrangements. In agreement
with~\cite{patinet_connecting_2016}, an excellent correlation is
observed. Above all, this correlation shows a slow decrease indicating
a persistence of weak sites.

We observe that the level of correlation depends on the preparation of the system. The more relaxed the system, the more robust the observed correlation. The first ten plastic rearrangements occur in areas belonging to the softest $23$, $13$ and $8.5\%$ sites for the HTL, ESL and GQ systems, respectively. For slowly quenched glasses GQ, it is interesting to note that the correlation decreases sharply for a deformation corresponding to the softening due to the localization of the deformation. It can be argued that the origin of the best correlation observed in the most relaxed system comes directly from the distribution of local thresholds. The relaxed systems have a much smaller population of low yield threshold zones. They therefore exhibit larger shear susceptibility when compared to other zones or to mechanical noise, making it easier to predict the onset of plastic activity.

The problem of the preceding method is that it assumes the existence
of individual and well-localized events. However,
Refs.~\cite{maloney_subextensive_2004,dasgupta_microscopic_2012} have
shown that if this hypothesis is relatively well satisfied for small
deformation levels, it does not hold with the increase in deformation
during which avalanches, through system spanning plastic events, are
observed. To circumvent this problem, we deal directly with the
correlation between the entire local yield stress field of the
as-quenched state $\Delta\tau_{y}$ and the cumulative deformation
field $\eta_{VM}$ in the same spirit as
Ref.~\cite{smessaert_structural_2014}. The cross-correlation, or
Pearson's correlation, is calculated as a function of the applied
strain $\gamma_{xy}$ as:
\begin{equation}
\label{eq:correlation_Pearson}
\rho_{\Delta\tau_{y},\eta_{VM}}(\gamma_{xy})=-\frac{\sum_{i=1}^{N}(\Delta\tau_{y}^{i}-\overline{\Delta\tau_{y}})(\eta_{VM}^{i}-\overline{\eta_{VM}})}{N\sigma_{\Delta\tau_{y}}\sigma_{\eta_{VM}}},
\end{equation}
where $N$ is the number of points on the grid on which the thresholds are calculated, $\sigma_{\Delta\tau_{y}}$ and $\sigma_{\eta_{VM}}$ are the standard deviations of $\Delta\tau_{y}$ and $\eta_{VM}$, respectively. The minus sign is added here to obtain a positive value since large $\eta_{VM}$ are expected for locations where $\Delta\tau_{y}$ are small (i.e. anti-correlation). $\overline{A}$ denotes the ensemble average of the quantity $A$. Note that explicit dependence on $\gamma_{xy}$ of $\eta_{VM}$ is omitted in the r.h.s for the sake of simplicity. Fig.~\ref{fig:correlations}b shows the evolution of $\rho_{\Delta\tau_{y},\eta_{VM}}$ as a function of the imposed deformation. The general trend is qualitatively similar to that of Fig.~\ref{fig:correlations}a. The correlation between local thresholds and plastic activity is greater for the more relaxed systems. There are, however, some differences. It can be observed that the correlation begins to increase as plastic rearrangements start to accumulate on weak sites. On the other hand, as expected, the decay of the GQ system is more marked upon the formation of shear bands, the latter concentrating the deformation.

The correlation appears smaller than the one expected from the computation based on local rearrangements in Eq. \ref{eq:correlation_CDF}, however, we remark that this calculation is based on crude assumptions. For example, we have not dissociated the elastic part from the plastic part when calculating $\eta_{VM}$. Moreover, this approach does not take into account the distribution of amplitudes of plastic rearrangements. Still, we have verified that in both cases - correlation based on the maxima of the strain field (Eq. \ref{eq:correlation_CDF}) and the cross-correlation based on the cumulative deformation (Eq. \ref{eq:correlation_Pearson}) - the correlations between $\Delta\tau_y$ and plastic activity are significantly better, and more persistent with deformation, than those obtained for the local classical structural indicators reviewed in~\cite{patinet_connecting_2016}.

\subsection{\label{sec:distributions_of_local_relaxation} Distributions of local relaxation}

In this section, we extend our method to study the amplitude of the
local plastic relaxations that follow plastic rearrangements.  The
loading of region-I described in Sec. \ref{sec:local_method} is
continued after the instability until the local stress $\tau(\alpha)$
increases again, signaling the end of the plastic rearrangement and
the return to mechanical stability of the sheared zone. This final
stress $\tau_{f}(\alpha) $ is also computed for all
directions. Plastic relaxation in the $\alpha$ direction is then
deduced by simply subtracting $\tau_{f}(\alpha)$ from the stress just
before the instability $\tau_{c}(\alpha)$. This amplitude of
relaxation $\Delta\tau_{r}(\alpha)=\tau_{c}(\alpha)-\tau_{f}(\alpha)$
is exemplified in Fig.~\ref{fig:local_method}b. Like the thresholds,
it depends on the direction of shear. Note that this method can
  only give an estimate of the relaxation amplitude, as the frozen
  boundary conditions constrain some degrees of freedom during
  relaxation.

\begin{figure}
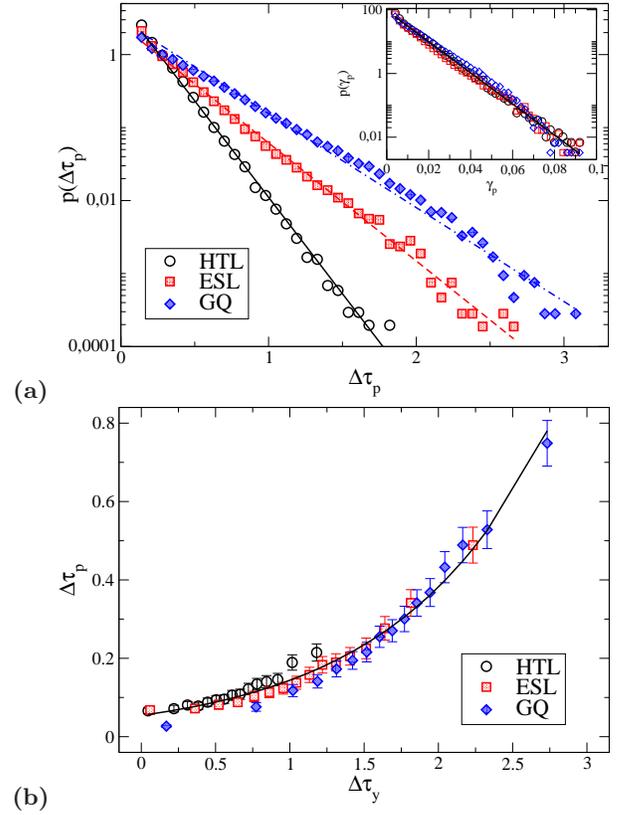

\textbf{(a)}
\includegraphics[width=0.85\columnwidth]{PDF_DeltaTaup_QP}\\
\textbf{(b)}
\includegraphics[width=0.85\columnwidth]{DeltaTaup_vs_DeltaTauy}
\caption{\label{fig:relaxations} a) Probability distribution function of the stress drops for the three different quench protocols in lin-log scale. The same quantities rescaled by the shear moduli, i.e. the slip increments, are represented in the inset. The lines are exponential fits. b) Average stress drop as a function of the average local yield stress for the free different quench protocols. The line is an exponential fit.}
\end{figure}

The plastic rearrangements actually observed during the shearing of the system correspond to the thresholds $\Delta\tau_{y}$ calculated in Eq. \ref{eq:loctauy}, and occur when the patch is loaded in the weakest direction $\alpha_{min}$. To place ourselves in a coarse-graining perspective, we want to derive a scalar indicator that corresponds to mechanical response in the remote loading direction $\alpha_l$ and disregard for now the tensorial aspect of the problem. The amplitude of plastic relaxation is therefore calculated, in turn, by projecting $\Delta\tau_{r}$ in the $\alpha_l$ direction according to the relation:
\begin{equation}
\label{eq:tau_p}
\Delta\tau_{p}=\Delta\tau_{r}(\alpha_{min})\cos(2[\alpha_{min}-\alpha_l]).
\end{equation}
Note that this estimator of the stress relaxation slightly underestimates the plastic relaxations because of the projection. Nevertheless, it gives access to a sufficiently simple scalar indicator. We verified that the absence of projection does not qualitatively change our results (not shown here).

The distributions of stress relaxation amplitudes reported in Fig.~\ref{fig:relaxations}a in lin-log scale for the three quench protocols show an exponential decay. The average stress drops increases with the relaxation of the system. The mean plastic relaxations calculated from exponential regressions are $\overline{\Delta\tau_{p}}=0.164$, $0.269$ and $0.337$ for HTL, ESL and GQ systems, respectively. The amplitude of plastic deformation, or slip increment, can also be estimated by computing the eigen-deformations of plastic rearrangements as $\gamma_{p}\sim\Delta\tau_{p}/\mu$ where $\mu$ is the average shear modulus of the glass. Remarkably, the distributions $\gamma_{p}$ collapse on a master curve of mean $\overline{\gamma_ {p}}=0.00887$ independently of the quench protocol as reported in the inset of Fig.~\ref{fig:relaxations}a. These results justify the assumption of a characteristic relaxation commonly used in mesoscopic simulations or in mean-field models. It is also in agreement with previous atomistic computations based on different methods such as the mapping between elastic field and Eshelby inclusion model~\cite{albaret_mapping_2016,boioli_shear_2017} and automatic saddle point search techniques~\cite{fan_energy_2017}.

We also take advantage of this analysis to study the dependence of the relaxation amplitude $\Delta\tau_{p}$ with the distance to thresholds $\Delta\tau_{y}$ as shown in Fig.~\ref{fig:relaxations}b. The former increases on average according to the latter. Remarkably, the relationship observed does not seem to depend too much on the quench protocol. If it seems reasonable that the amplitude of relaxation increases with the increase of the local yield stress, the stored elastic energy being larger, we have no explanation to derive this relationship at the moment. An exponential dependence is adjusted empirically and gives $\Delta\tau_{p}\sim0.054e^{0.976\Delta\tau_{y}}$.

Let us note finally that if we are sufficiently confident in the capacity of this local method to quantify the thresholds, the measurements of the relaxation amplitudes are more questionable insofar as the frozen boundary conditions prevent some relaxations. A more adequate treatment of this question would require developments that are beyond the scope of this work. One may for instance think about the implementation of quasicontinuum simulation techniques that, by relaxing elastically the surrounding matrix, will provide flexible boundary conditions to the atomistic region. The picture that emerges from Fig.~\ref{fig:relaxations} is nevertheless interesting and sheds new light on the plastic deformation as  it greatly simplifies representation of relaxation in glassy systems.

\section{\label{sec:orientation_effects} Orientation effects}

\subsection{\label{sec:loading_direction} Loading direction}

So far, relatively few studies have addressed the issue of the variation of the local yield stress as a function of loading orientation. This is due to the fact that most of the proposed local indicators are scalar quantities. Only work based on soft modes attempted to explore susceptibility to loading orientation by taking advantage of the vectorial aspect of vibrational eigenmodes~\cite{rottler_predicting_2014,smessaert_structural_2014}. In order to address this issue, we use here another asset of our local method which naturally gives us access to this directional information.

\begin{figure*}
\includegraphics[width=1.9\columnwidth]{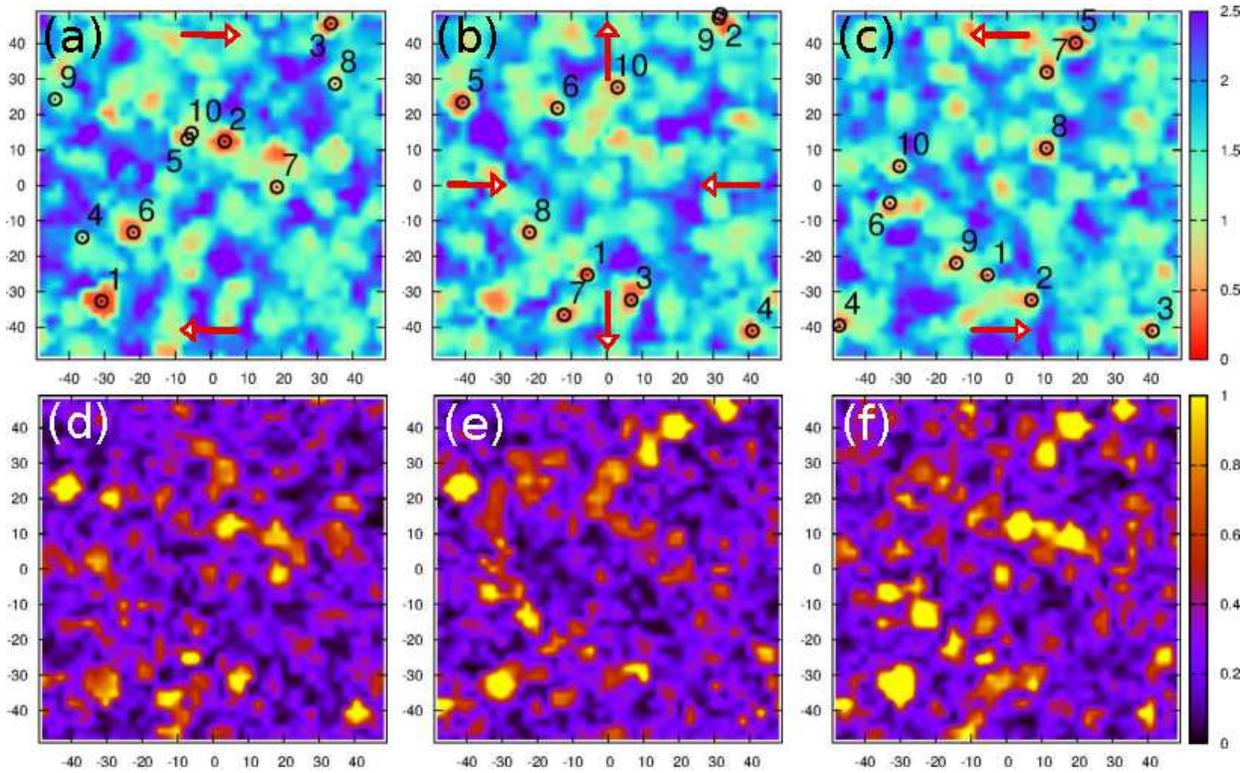}
\caption{\label{fig:yield_stress_map_orientation_effect} Top row: local yield stress maps computed on a regular grid for different loading directions for the GQ protocol: a) simple shear $\alpha_{l}=0^{\circ}$, b) pure shear $\alpha_{l}=45^{\circ}$ and c) negative simple shear $\alpha_{l}=90^{\circ}$. The red arrows correspond to the applied strains. The first ten plastic event locations are shown as open black symbols numbered by order of appearance during remote shear loading. Bottom row: local yield stress contrasts $TC(\alpha_{l}^1,\alpha_{l}^2)$ defined in Eq. \ref{eq:threshold_contrast} between the above loading directions: d) $\alpha_{l}^1=0^{\circ}$, $\alpha_{l}^2=45^{\circ}$ ,e) $\alpha_{l}^1=45^{\circ}$, $\alpha_{l}^2=90^{\circ}$ and f) $\alpha_{l}^1=0^{\circ}$, $\alpha_{l}^2=90^{\circ}$.}
\end{figure*}

To test the dependence on the direction of the mechanical loading, the quenched glasses are deformed by the same AQS protocol but following different orientations. In addition to the simple shear described in Sec. \ref{sec:loading}, the systems are deformed in pure shear by applying strain increments $\Delta\gamma/2 =-\Delta\epsilon_{xx}=\Delta\epsilon_{yy}$ and simple shear in the negative direction by applying deformation increments $\Delta\gamma =-\Delta\gamma_{xy}$. These remote loadings correspond in the infinitesimal strain limit to shearing along $\alpha_{l}=45^{\circ}$ and $\alpha_{l}=-90^{\circ}$ directions, respectively. Pure shear thus produces a diagonally-oriented shear. The negative simple shear corresponds to a laterally-oriented shear but in the opposite direction with respect to the positive simple shear remote loading employed so far. The positions of the first ten plastic rearrangements for the different loading directions are exemplified for a GQ glass in the top row of Fig.~\ref{fig:yield_stress_map_orientation_effect}. In agreement with Ref.~\cite{gendelman_shear_2015}, the location of plastic rearrangements show a strong dependence on the loading protocol. Most plastic events occur in different areas for different loading protocols. Only occasionally rearrangements will appear in the same location.

At the same time, local yield stresses are also calculated using the formula \ref{eq:loctauy} with the corresponding loading directions $\alpha_{l}$. $\alpha_{l}$ is thus equal to $0^{\circ}$, $45^{\circ}$ and $90^{\circ}$ for positive simple shear, pure shear and negative simple shear, receptively. The maps of $\Delta\tau_{y}$ are shown in the top row of Fig.~\ref{fig:yield_stress_map_orientation_effect}. A strong dependence on the loading orientation is observed. The rotation of the shear results in the appearance (disappearance) of soft (hard) zones. For example, the areas close to the 1st and 3rd plastic rearrangements for positive simple shear in Fig.~\ref{fig:yield_stress_map_orientation_effect}a disappear in the case of negative simple shear in Fig.~\ref{fig:yield_stress_map_orientation_effect}c. Conversely, soft areas appear as those close to the 5th and 7th rearrangements in Fig.~\ref{fig:yield_stress_map_orientation_effect}c. As with the simple shear detailed above, pure shear and negative simple shear show an excellent correlation between the soft zones and the zones where the plastic rearrangements occur. The quantification of correlations through Eq.~(\ref{eq:correlation_CDF}) and (\ref{eq:correlation_Pearson}) as described in Sec.~\ref{sec:correlation_with_plastic_activity} is quantitatively similar (not shown here).

\begin{figure}
\includegraphics[width=0.85\columnwidth]{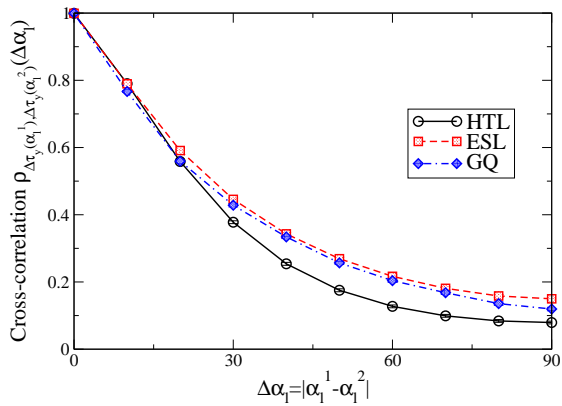}
\caption{\label{fig:Cross_Correlation_QP_vs_AlphaL} Cross-correlation of the local yield stress field as a function of the loading direction shift. Error bars are smaller than symbols.}
\end{figure}

In order to highlight the discrete aspect of the variation of the local yield stress field, we compute the threshold contrast $TC$ existing between two loading directions. This contrast is defined locally as the ratio between their difference and their averages as
\begin{equation}
\label{eq:threshold_contrast}
TC(\alpha_{l}^1,\alpha_{l}^2)=\frac{|\Delta\tau_y(\alpha_{l}^1)-\Delta\tau_y(\alpha_{l}^2)|}{(\Delta\tau_y(\alpha_{l}^1)+\Delta\tau_y(\alpha_{l}^2))/2},
\end{equation}
where $\alpha_{l}^1$ and $\alpha_{l}^2$ are two remote loading directions. Contrast maps are shown in the bottom row of Fig.~\ref{fig:yield_stress_map_orientation_effect}. These maps feature the trends described qualitatively above. The change in loading angle clearly shows areas of marked contrasts as a function of the loading orientations considered. The change of loading direction ``turns on" or ``turns off" the soft areas which gives rise to large local contrasts.

We observe that the greater the difference between the angles $\Delta\alpha_{l}=|\alpha_{l}^1-\alpha_{l}^2|$, the greater the number and intensity of the contrasts. In order to quantify this trend, we compute the cross-correlation of the yield stress field as a function of the difference of the loading angles $\Delta\alpha_{l}$. The result is shown in Fig.~\ref{fig:Cross_Correlation_QP_vs_AlphaL}. The trend observed confirms that the correlation of the yield stress field decreases rapidly with the loading angle, regardless of the quench protocol. We note, however, that the correlation is never completely zero, and is still significant even for the largest $\Delta\alpha_{l}=90^{\circ}$ that corresponds to the correlation between a shearing direction and its opposite direction. We attribute this effect to the small correlation existing between stable (unstable) zones and their tendency to have large (small) slip barriers~\cite{rodney_yield_2009,patinet_connecting_2016}. The decorrelation due to the directional aspect is nevertheless clearly the dominant effect.

This result shows that local stress thresholds are a very sensitive probe of the loading protocol insofar as it is possible to acurately predict the plastic activity as a function of the orientation of the load. Unlike~\cite{gendelman_shear_2015}, we thus believe that these results are consistent with a plasticity-based view of shear transformation zones. Indeed, from our point of view, the dependence of the plastic activity upon the loading protocol does not rule out the existence of plastic deformation via discrete units encoded in the structure, and that these discrete units clearly preexist within the material prior to loading. Our results show rather that the plastic deformations of an amorphous solid, at least for the transient regime at small deformations, can be seen as a sequence of activation of discrete shear transformation zones having weak slip orientations.

\subsection{\label{sec:loading_direction} Fluctuations}

\begin{figure}
\includegraphics[width=0.85\columnwidth]{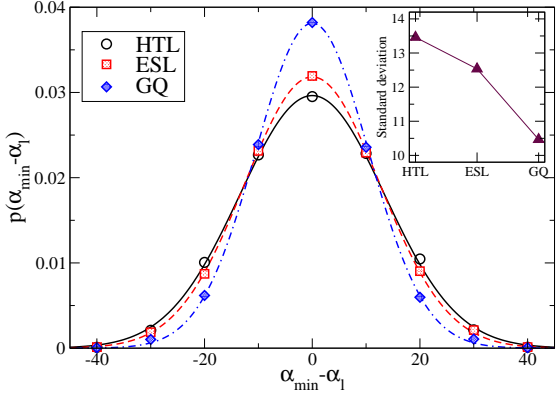}
\caption{\label{fig:orientation_effect} Probability distribution functions of angles for the three quench protocols for which the projected local yield stress is minimal $\alpha_{min}$. The lines correspond to Gaussian fits which standard deviations are reported in the inset.}
\end{figure}

The local information given by our method allows us to also study the
fluctuations in the direction of plastic rearrangements around the
loading direction $\alpha_{l}$. At first, we have verified that the
distributions of thresholds do not depend on the angle
$\alpha_{l}$. As expected, the glasses are isotropic on average in the
as-quenched state. We then consider the angle $\alpha_{min}$
minimizing Eq.~(\ref{eq:loctauy}) for a given $\alpha_{l}$, i.e. the
weakest local direction for a given loading direction. The
distributions of $\alpha_{min}$ around $\alpha_{l}$, shown in figure
\ref{fig:orientation_effect}a, are well described by a Gaussian
function of standard deviation $\sigma_{\alpha_{min}} \sim
12^{\circ}$. The latter decreases slightly with the relaxation of the
system.

\begin{figure}
\includegraphics[width=0.85\columnwidth]{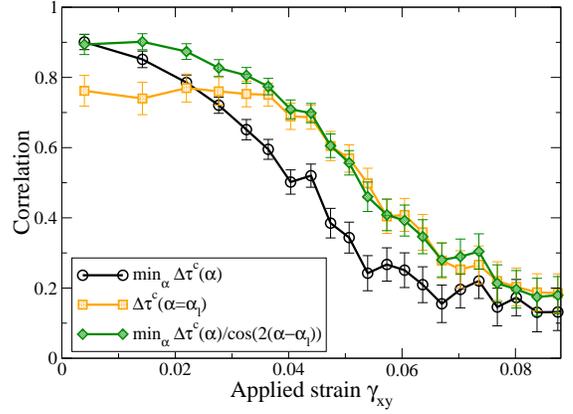}
\caption{\label{fig:correlation_orientation_effect} Same as Fig.~\ref{fig:correlations}a for the GQ protocol for different local yield stress fields: its minimum over all orientations $\min_{\alpha}\Delta\tau_{c}(\alpha)$, its value along the loading direction $\Delta\tau_{c}(\alpha=\alpha_{l})$ and its minimum once projected along the loading direction $\min_{\alpha}\Delta\tau_{c}(\alpha)/\cos(2(\alpha-\alpha_{l}))$ (as in Fig.~\ref{fig:correlations}a).}
\end{figure}

We also examine the consequences of taking into account the different possibilities of rearrangement directions on the correlation between local yield stress and plastic activity. Three types of local indicators can be considered: the minimum $\Delta\tau_{c}(\alpha)$ over all directions, the threshold $\Delta\tau_{c}(\alpha=\alpha_{l})$ only along the loading direction and $\Delta\tau_{y}$ as previously defined in Eq. \ref{eq:loctauy}. The correlations calculated for these three quantities from the relation \ref{eq:correlation_CDF} are plotted in Fig.~\ref{fig:correlation_orientation_effect} as a function of the imposed deformation.
We observe that the minimum over all angles $\min_{\alpha}\Delta\tau_{c}(\alpha)$ gives the best correlations only for the very first plastic rearrangements and then decreases rapidly with deformation. This indicator corresponds to isotropic excitation, such as fluctuations in thermal energy, and is therefore sensitive to small barriers. Conversely, $\Delta\tau_{c}(\alpha=\alpha_{l})$ shows a poorer correlation with the location of plastic activity for the first rearrangements as it misses the low thresholds which are slightly disoriented with respect to the loading direction of the system. On the other hand, the correlation is better for larger deformations. Finally, $\Delta\tau_{y}$ shows the best correlation for both small and large deformations. Due to the projection, it is sensitive to small thresholds while retaining information specialized for macroscopic loading direction for larger yield stresses. We see here the importance of having access to a directional quantity. The local yield stresses defined in this article are therefore a good compromise between simplicity (purely local and scalar) and performance that justifies our approach. Note that qualitatively similar results have been obtained as a function of the relaxation of the system or when the correlations are computed from Eq.~(\ref{eq:correlation_Pearson}) (not shown here).

\section{\label{sec:length_scale} Length scale of the local probing zone}

\subsection{\label{sec:optimal_size} Optimal size}

\begin{figure}
\includegraphics[width=0.85\columnwidth]{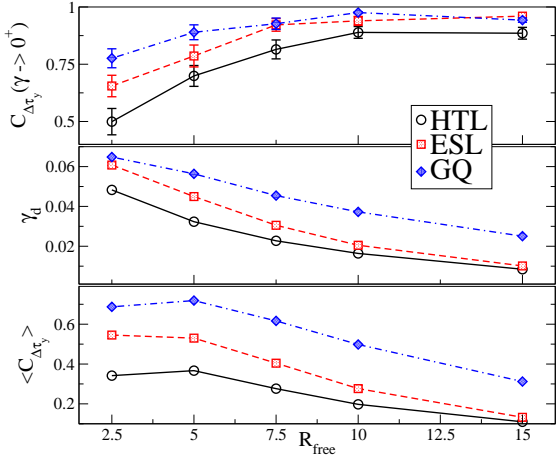}
\caption{\label{fig:correlation_size_effect} Correlation indicators computed as a function of the size of the probing zone $R_{free}$. Top: Correlation with the first plastic rearrangement locations. Middle: decorrelation strain. Bottom: Average correlation.}
\end{figure}

We are interested here in the effects of the patch size $R_{free}$
(Region I) on which the local yield stresses are computed. $R_{free}$
is varied from $2.5$ to $15 \sigma$. The procedure is the same as
described in Sec.~\ref{sec:local_method}. The size of the grid
$R_{sampling}$ on which $\Delta\tau_{y}(R_{free})$ is sampled is kept
constant and equal to $L/39 \approx R_{cut}$. We first investigate the
correlations of the thresholds with the plastic activity using
relation~(\ref{eq:correlation_CDF}). To quantify the degree of
correlation, three kinds of indicators are considered: the correlation
of the first plastic rearrangements
$C_{\Delta\tau_{y}}(R_{free},\gamma_{xy}\rightarrow 0^{+})$, the
characteristic deformation $\gamma_{d}$ on which the correlation
decreases with imposed deformation and the average correlation over
the investigated strain window $\langle
C_{\Delta\tau_{y}}(R_{free})\rangle$. The variations of these three
indicators as a function of $R_{free}$ are shown for the three quench
protocols in Fig.~\ref{fig:correlation_size_effect}.

\begin{figure*}
\includegraphics[width=1.9\columnwidth]{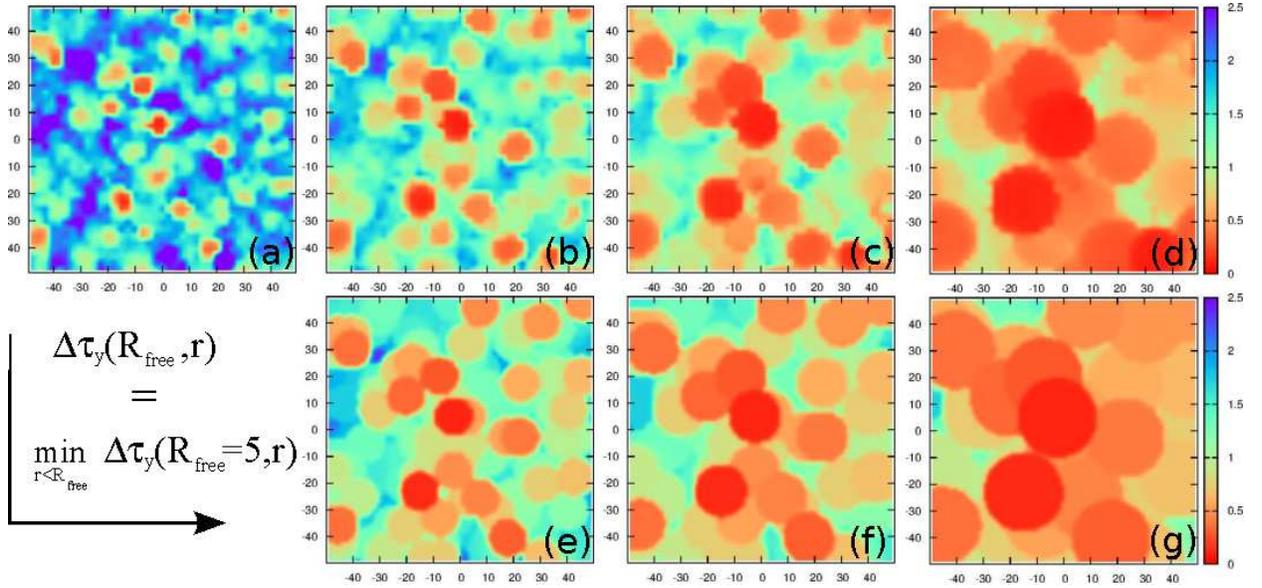}
\caption{\label{fig:yield_stress_map_size_effect} Top row: Local yield stress maps of a GQ glass computed for different inclusion sizes $R_{free}$ : a) $5$, b) $7.5$, c) $10$ and d) $15$. Bottom row: Corresponding local yield stress maps deduced from local minima obtained from the map computed for $R_{free}=5$ shown in figure a.}
\end{figure*}

\begin{figure*}
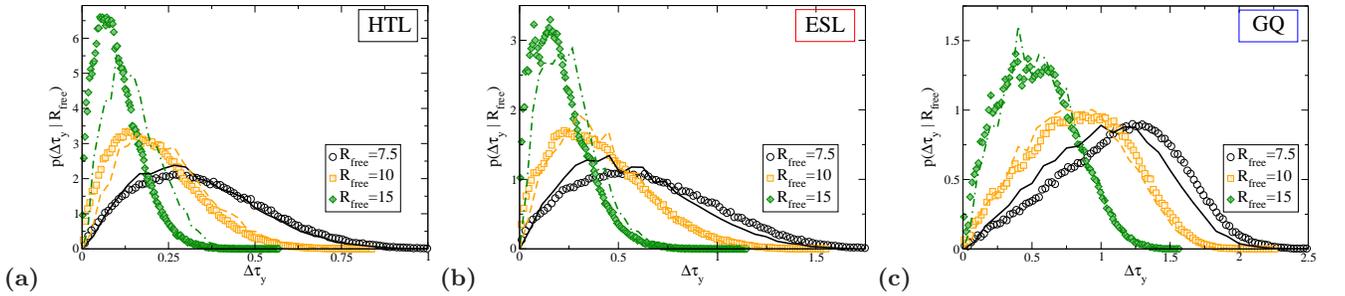

\textbf{(a)}
\includegraphics[width=0.59\columnwidth]{PDF_DeltaTauy_vs_Rfree_HTL}
\textbf{(b)}
\includegraphics[width=0.59\columnwidth]{PDF_DeltaTauy_vs_Rfree_ESL}
\textbf{(c)}
\includegraphics[width=0.61\columnwidth]{PDF_DeltaTauy_vs_Rfree_GQ}
\caption{\label{fig:PDF_DeltaTauy_vs_Rfree} Probability distribution function of the local yield stress for the three different quench protocols as a function of the inclusion size $R_{free}$: a) HTL, b) ESL and c) GQ. The lines correspond to the zoom-out process exemplified in the bottom row of Fig.~\ref{fig:yield_stress_map_size_effect} where the local yield stresses are deduced from maps computed with $R_{free}=5$.}
\end{figure*}

The correlation of the first plastic rearrangement $C_{\Delta\tau_{y}}(R_{free},\gamma_{xy}\rightarrow 0^{+})$ increases with the size $R_{free}$. Indeed, the increase of the probing zone makes it possible to progressively integrate the elastic loading heterogeneities. The loading felt by the sheared zones converges with $R_{free}$ toward the effective loading produced by a remote loading, which makes it easier to identify the weak zones. For $R_{free}<5$, we observe a marked drop of the correlation. For this small size, in addition to larger elastic loading heterogeneities, the frozen boundary conditions over-constrain the measurement of local shear stress thresholds.

The decorrelation strain $\gamma_{d}$ is extracted from a fit of the
curves $C_{\Delta\tau_{y}}(R_{free},\gamma_{xy})$ with the
  expression: $A+Be^{-(\gamma_{xy}/\gamma_{d})^2}$ as shown in
Fig.~\ref{fig:correlations}a. This deformation, corresponding to the
characteristic deformation on which the glasses lose their memory of
the quenched state, decreases with $R_{free}$. Indeed, after the first
plastic events, the use of a large $R_{free}$ loses information about
the hard zones surrounding the softest ones. A small $R_{free}$ allows
us, while still having good spatial resolution, to maintain a
significant correlation for higher deformations, the hard zones being
simply advected during plastic flow.

The last kind of correlation indicator is the average of $C_{\Delta\tau_{y}}$ computed as:
\begin{equation}
\langle C_{\Delta\tau_{y}}(R_{free})\rangle=(1/\gamma_{*})\int_{0}^{\gamma_{*}}C_{\Delta\tau_{y}}(R_{free},\gamma_{xy})d\gamma_{xy}.  
\end{equation}
The upper bound of the interval of integration $\gamma_{*}$ is chosen equal to the largest decorrelation strain $\gamma_{*}=\gamma_{d}(R_{free}=2.5)$, i.e. computed for the smallest $R_{free}$. This is a global indicator that gathers information on the degree of correlation at the origin and during deformation as the glass loses its memory from the quench state. The results reported in Fig.~\ref{fig:correlation_size_effect} show an overall decrease of the average correlation with $R_ {free}$.  This decrease is less marked between $R_{free}=2.5$ and $R_{free}=5$. The maximum of average correlation is even found for $R_{free}=5$ for the quench protocols HTL and GQ.

These results show empirically that a patch size of $R_{free}=5$ is a good compromise in terms of correlation between $\Delta\tau_{y}$ and plastic activity. Calculating the stress thresholds over this scale allows one to precisely locate the first plastic events while preserving the spatial resolution and keeping the memory of the initial quenched state. The effect of quench protocols is qualitatively similar to our previous observations. A greater relaxation of the system results in both a greater correlation for the first plastic events as well as a larger characteristic decorrelation strain, resulting in a larger average correlation.

\subsection{\label{sec:statistical_size_effects} Statistical size effects}

We are interested here in the effect of the patch size $R_{free}$ on
the slip barrier statistic.
Several mechanisms such as mechanical and statistical size effects
  can be anticipated. Mechanical size effects correspond to elastic
  heterogeneities as well as the influence of frozen boundary
  conditions. Frozen boundaries affect the simulation in the following
  way: the closer an atom to the boundary, the more affine its
  displacement, thus deviating its trajectory with respect to non
  constrained simulations. Statistical size effects play a role insofar as the local yield stress is primarily controlled by the weakest
  zones in the patch since its amplitude is given by the smallest
  threshold contained in the patch. Maps of local $\Delta\tau_{y}$
computed for different $R_{free}$ are shown in
Fig.~\ref{fig:yield_stress_map_size_effect} (top row) for the quench
protocol GQ. We observe that the variation of $R_{free}$ modifies the
global statistic of the thresholds. The distribution functions
presented in Fig.~\ref{fig:PDF_DeltaTauy_vs_Rfree} for the three
quench protocols show that the increase of $R_{free}$ induces a
significant shift of the distributions toward smallest values of
$\Delta\tau_{ y}$.

Obviously, the maps obtained for the large $R_{free}$ can be explained by the spatial increase of the zones centered on weak sites. The softest areas tend to ``invade'' the glass as the radius of the area on which the threshold is computed increases. Hence, the statistical effect seems to be dominant. On the basis of this observation, we try to understand the variations of the distributions of the local yield stresses with $R_{free}$. We choose to work from the observed distributions for a size $R_{free}=5$. We make the simplifying hypothesis that all the thresholds $\Delta\tau_{y}(R_{free})$ of the grid points take the value of the smallest local minima $\Delta\tau_{y}(R_{free}=5)$ located inside a disk of radius $R_{free}$. For comparison, maps deduced by this procedure are given in Fig.~\ref{fig:yield_stress_map_size_effect} (bottom row). This purely geometric approach shows a remarkable agreement compared to the local yield stress maps calculated by actually varying $R_{free}$.

This approach allows us to deduce the distribution of the yield stresses as a function of a given patch size $R_{free}>5$ from the distribution obtained for $R_{free}=5$. The comparisons between the distributions computed for the three quench protocols for different $R_{free}$ and those estimated from our procedure reported in Fig.~\ref{fig:PDF_DeltaTauy_vs_Rfree} show a satisfactory agreement. The variation of the distributions of $\Delta\tau_{y}$ is therefore dominated by statistical effects. The increase of $R_{free}$ plays the role of a low-pass filter for the thresholds, shifting their distributions toward smaller yield stress values. The agreement between the measured distributions and the deduced distributions is nevertheless slightly lower for the less relaxed systems and for the large $R_{free}$ values. We attribute this discrepancy to the larger elastic disorder and to the lower sensitivity of the soft zones due to narrower threshold distributions in these systems.

\section{\label{sec:conclusion} Conclusions}

In this article, we describe a method for sampling local slip thresholds in model amorphous solids. A robust correlation is observed between the zones with small yield stresses and the locations where the plastic rearrangements occur. As expected, the more the state of the glass is relaxed, the more the barrier distributions shift towards the larger values, explaining the strengthening of glasses from their local stability. 

This local method has been extended to measure the amplitude of the plastic relaxations. We show that the assumption of a characteristic mean plastic relaxation is reasonable, the relaxation amplitudes following exponential distributions. Interestingly, we have shown that the amplitude of the plastic relaxations increases on average with the yield stresses. 

The effects of loading orientation have shown that the variation of the plastic activity with the direction of loading is well captured by the variation of the local yield stress field calculated using our method. Finally, the variation of the threshold statistics with the size of the probing zones can be reproduced with reasonable agreement on the basis of simple geometric arguments. These results reinforce the coherence of the amorphous plasticity modeling based on discrete flow defects that possess weak slip directions and which are encoded in the structure of the material.

The advantages of the method presented in this work are numerous. It allows to probe the local slip thresholds in a non-perturbative way over a well-defined length scale. Moreover, its generalization to other atomic systems does not seem to pose any particular difficulty since it is, in principle, transposable to all glassy solids. Finally, a last advantage of this method is its computational cost. While, for instance, normal mode analysis based-methods scale as the cube of the number of atoms, our method scales linearly with it. Furthermore, as it treats the different parts of the solid independently it is by construction suited for massively parallel simulations. It is therefore possible to handle extended systems.

The implementation of this local method opens up several promising perspectives. It will be interesting to compare quantitatively the predictive power of the plastic activity of this method with other recent works also providing robust indicators of plasticity~\cite{rodney_distribution_2009,rodney_yield_2009,fan_how_2014,fan_crossover_2015, fan_energy_2017,ding_universal_2016, cubuk_identifying_2015, schoenholz_structural_2016,cubuk_structural_2016}.

Future research could focus on the measurement of quantities on an atomic scale needed for coarse-grained approaches~\cite{hinkle_coarse_2017}. For instance, our method  can provide the threshold statistics necessary to take into account the disorder in the mesoscopic models~\cite{BulatovArgon94a,BVR-PRL02,Picard-PRE02,vandembroucq_mechanical_2011,TPVR-Meso12,Nicolas-SM14,tyukodi_depinning_2016,nicolas_deformation_2017} and could explicitly deal with the tensorial nature of the problem and the effect of the loading geometry~\cite{budrikis_universal_2017}. 

This work paves the way, for example, to study the correlation between local energy barriers and frequencies of thermally activated rearrangements simulated by molecular dynamics. Hence, an important question left for future work is to study the effect of thermomechanical history on the statistics of local yield stresses. Our method will allow us to test some of the many phenomenological hypotheses upon which continuum models are still based~\cite{rottler_unified_2005,falk_dynamics_1998,sollich_rheology_1997,hebraud_mode-coupling_1998} and thus significantly improve the multi-scale modeling of plasticity of amorphous solids.

\begin{acknowledgments}
M.L. and S.P. acknowledge the support of French National Research Agency through the JCJC project PAMPAS under grant ANR-17-CE30-0019-01. R.G.G. and A.H.G acknowledge the support of CNRS and French National Research Agency under grant ANR-16-CE30-0022-03. M.L.F acknowledges the support of the U.S. National Science Foundation under Grant Nos. DMR1408685/1409560.
\end{acknowledgments}

\bibliography{bilbio_BLHGFVP}

\end{document}